**Vaccination strategies to control Ebola epidemics in the context of variable household inaccessibility levels**

G. Chowell[*,1,2], A. Tariq[1], M. Kiskowski[3]

[1]Department of Population Health Sciences, School of Public Health, Georgia State University, 30303, Atlanta, GA, USA

[2] Division of International Epidemiology and Population Studies, Fogarty International Center, National Institutes of Health, Bethesda, MD, USA.

[3] Department of Mathematics and Statistics, University South Alabama, 36688, Mobile, AL, USA

[*]Corresponding author

**Abstract**

Despite a very effective vaccine, active conflict and community distrust during the ongoing DRC Ebola epidemic are undermining control efforts, including a ring vaccination strategy that requires the prompt immunization of close contacts of infected individuals. However, in April 2019, it was reported 20% or more of close contacts cannot be reached or refuse vaccination [1], and it is predicted that the ring vaccination strategy would not be effective with such a high level of inaccessibility [2]. The vaccination strategy is now incorporating a "third ring" community-level vaccination that targets members of communities even if they are not known contacts of Ebola cases. To assess the impact of vaccination strategies for controlling Ebola epidemics in the context of variable levels of community accessibility, we employed an individual-level stochastic transmission model that incorporates four sources of heterogeneity: a proportion of the population is inaccessible for contact tracing and vaccination due to lack of confidence in interventions or geographic inaccessibility, two levels of population mixing resembling household and community transmission, two types of vaccine doses with different time periods until immunity, and transmission rates that depend on spatial distance. Our results indicate that a ring vaccination strategy alone would not be effective for containing the epidemic in the context of significant delays to vaccinating contacts even for low levels of household inaccessibility and affirm the positive impact of a supplemental community vaccination strategy. Our key results are that as levels of inaccessibility increase, there is a qualitative change in the effectiveness of the vaccination strategy. For higher levels of vaccine access, the probability that the epidemic will end steadily increases over time, even if probabilities are lower than they would be otherwise with full community participation. For levels of vaccine access that are too low, however, the vaccination strategies are not expected to be successful in ending the epidemic even though they help lower incidence levels, which saves lives, and makes the epidemic easier to contain and reduces spread to other communities. This qualitative change occurs for both types of vaccination strategies: ring vaccination is effective for containing an outbreak until the levels of inaccessibility exceeds approximately 10% in the context of significant delays to vaccinating contacts, a combined ring and community vaccination strategy is effective until the levels of inaccessibility exceeds approximately 50%. More broadly, our results underscore the need to enhance community engagement to public health interventions in order to enhance the effectiveness of control interventions to ensure outbreak containment.

**Author summary**

In the context of the ongoing Ebola epidemic in DRC, active conflict and community distrust are undermining control efforts, including vaccination strategies. In this paper, we employed an individual-level stochastic structured transmission model to assess the impact of vaccination strategies on epidemic control in the context of variable levels of household inaccessibility. We found that a ring vaccination strategy of close contacts would not be effective for containing the epidemic in the context of significant delays to vaccinating contacts even for low levels of household inaccessibility and evaluate the impact of a supplemental community vaccination strategy. For lower levels of inaccessibility, the probability of epidemic containment increases over time. For higher levels of inaccessibility, even the combined ring and community vaccination strategies are not expected to contain the epidemic even though they help lower incidence levels, which saves lives, makes the epidemic easier to contain and reduces spread to other communities. We found that ring vaccination is effective for containing an outbreak until the levels of inaccessibility exceeds approximately 10%, a combined ring and community vaccination strategy is effective until the levels of inaccessibility exceeds approximately 50%. Our findings underscore the need to enhance community engagement to public health interventions.

**Introduction**

The ongoing Ebola outbreak in the Democratic Republic of Congo (DRC), which has been active in the region for over a year (August 2018 to August 2019), has become the most complex Ebola epidemic to date, threatening to spread to neighboring countries.  Active case finding and contact tracing activities played a major role in controlling the 2014-2016 Ebola epidemic, which devastated communities in West Africa with a total of >28,000 cases and >11,000 deaths reported in the three most affected countries, Guinea, Liberia and Sierra Leone [3]. Although contact tracing is a critical piece of a response to Ebola outbreaks, its effectiveness varied over time across all three of the most affected countries [4-6]. Factors that hampered the effectiveness of contact tracing during the West African epidemic included geographic inaccessibility and socio-cultural challenges such as mistrust of healthcare workers and community resistance to case investigation and contact tracers [6-8].

The unprecedented epidemic in DRC is unfolding in a climate of community distrust in the interventions [9, 10] in an active conflict zone [11], where over 200 attacks have deliberately targeted healthcare workers and treatment centers involved in the Ebola response efforts, allowing sustained transmission in the region. Consequently, the DRC Ebola epidemic has accumulated 2763 cases including 1841 deaths as of August 4, 2019 and has reached the urban city of Goma [12] and spilled over to Uganda where 3 cases have been confirmed as of June 12, 2019 [13, 14]. A surge in case incidence during the last few months (April-August, 2019) has coincided with an increasing trend in the number of violent attacks on health centers and health teams fighting the epidemic in the region [15].

Although control efforts now employ a highly effective emergency vaccine, the ongoing Ebola epidemic in the DRC is the first to occur in an active conflict zone. Deliberate violent attacks and threats to health workers on a scale not seen in previous Ebola outbreaks are directly undermining active control efforts in the region [11]. A ring vaccination strategy [16, 17] that was highly effective in the capital [18] was introduced rapidly on August $8^{th}$, 2018 within 8 days of the outbreak declaration in the provinces of North Kivu and Ituri [19]. However, it has been challenging to implement an effective ring vaccination strategy in a climate of violence targeting healthcare workers and decreasing accessibility to a mobile, fractionated population as it relies on the identification of contacts and contacts of contacts [20]. This issue is compounded by a large fraction of Ebola cases unconnected to known chains of transmission [21]. In order to boost the impact of the vaccine in a challenging transmission setting with a large fraction of hidden or underground transmission, the vaccination strategy is now incorporating vaccination of third level contacts [22-24] (also known as "third ring" or "ring +" vaccination). This is a community-level vaccination because the third ring strategy will immunize members of communities even if they are not known contacts of Ebola cases[24] and increases the number of contacts vaccinated from as many as 120 to 210 [25]. Also being considered is using a lower vaccine dose for the third ring vaccination, which will take longer to elicit protective antibody levels in vaccinated individuals [22, 23].

For simplicity, the effects of suboptimal control interventions including contact tracing and vaccination at the household level can be investigated using a structured individual-level model consisting of two spatial scales: households and overlapping communities [26, 27]. Households are then categorized in two types: accessible and inaccessible households whereby individuals in *inaccessible households* do not provide contacts lists to contact tracing teams, which hinders the effectiveness of contact tracing efforts, and do not participate in vaccination. This lack of participation in contact tracing may be due to a range of factors including health care inaccessibility, and furthermore, vaccination refusal may occur when contacts have religious beliefs that do not permit them to take the Ebola vaccine, they may think they do not need it or they may not believe in Ebola [24, 28].

Here we sought to evaluate the impact of ongoing Ebola vaccination campaigns by extending an individual-based model with household-community mixing [26, 27], which has been previously used to analyze the effects of lower versus higher rates of population mixing on Ebola transmission dynamics. This model has been able to fit the growth patterns of the 2014-16 Western Africa Ebola epidemic (e.g., the growth pattern of the epidemic in Guinea, versus Sierra Leone or Liberia) by calibrating the extent of community mixing [26]. This structured transmission model predicts outbreaks that propagate through the population as spatial waves with an endemic state [27] and has been useful to gain insight on the level of control that would be required to contain Ebola epidemics [27]. One of the less intuitive results of the community model was that even a low daily incidence can indicate an epidemic that is difficult to extinguish if saturation effects decrease incidence and are masking a higher reproductive number. Saturation effects occur when contacts of infectious individuals are already infected by other members of the community, such as family members or members of other close groups, and decrease the incidence when community mixing is low.

Assessing the effect of targeted vaccination efforts requires mathematical models that capture the contact structure of the community network [29-33]. In our paper, we aimed to investigate the effect of ring [29, 31, 34-36] and community vaccination strategies on outbreak control. For this purpose, our baseline stochastic model, which has been calibrated using data for the Western African Ebola epidemic[26, 27], was adapted to incorporate key features of the Ebola epidemic in the DRC. Specifically, epidemiological and transmission parameters, such as the incubation and infectious periods, and the household and community reproductive numbers, were based on the West African Ebola epidemic in general. The community size, which corresponds to the connectivity of the contact network, was based on data from the Liberian epidemic for the structure of a contact network in an urban setting.[1] This baseline model was adapted for two different programs of vaccination and two types of vaccine doses, a variable fraction of the population that is accessible and distance-dependent transmission rates. With the updated model, we evaluate the effectiveness of control strategies in the context of a fraction of the population that is accessible to vaccination teams, tied to the success of contact tracing and vaccination efforts. Our baseline network transmission model also adapts for the ring vaccination strategy by integrating heterogeneity in community transmission rates that scale with the distance between an infectious individual and each member of that individual's community. As the distance between an infectious individual and a contact increases, the transmission rate decreases exponentially (or by any other function). Ring and community vaccination strategies are compared by accounting for the two different spatial scales for the radius of contacts that are vaccinated (either as a ring or community-wide) and the two different vaccine doses (a full or half dose), that result in different time periods for the immunization of vaccinated contacts.

---

[1] Previously, the different growth profiles of the Ebola Virus epidemics in Guinea, Sierra Leone and Liberia were matched by varying the network structure (community size C) rather than modifying intrinsic transmission rates 26.     Kiskowski M. Three-Scale Network Model for the Early Growth Dynamics of 2014 West Africa Ebola Epidemic. . PLOS Currents Outbreaks. 2014;doi: 10.1371/currents.outbreaks.b4690859d91684da963dc40e00f3da81. doi: doi: 10.1371/currents.outbreaks.c6efe8274dc55274f05cbcb62bbe6070, 27.     Kiskowski M, Chowell G. Modeling household and community transmission of Ebola virus disease: epidemic growth, spatial dynamics and insights for epidemic control. Virulence. 2015;7(2):63-73. doi: doi: 10.1080/21505594.2015.1076613. PubMed PMID: 26399855.

**Methods**

**Household-Community Model for Ebola Spread and Two Types of Vaccination with Accessible and Inaccessible Households**

We previously described our model for disease spread [26, 27] and vaccination [37] in a contact network with household-community structure. A new component of our model is that we add a second type of household that is inaccessible to vaccination teams. These households may not participate for a variety of reasons including Ebola cases that are not identified, lack of confidence in the vaccination program and geographic inaccessibility. The inaccessible households do not participate in vaccination and do not provide contacts lists. We also modify the model to include two vaccine types. A larger vaccine dose is used for ring vaccination with faster immunization occurring 10 days later, and a smaller vaccine dose is used for community vaccination with immunization occurring 28 days later. This corresponds to the dosing and predicted timing of the ring and community vaccination regimens of 0.5 ml and 0.2 ml doses, respectively, recently described for the DRC Ebola epidemic [20, 22].

**Disease and Vaccination States**

The progression of Ebola disease, vaccination and immunization are modeled with seven epidemiological states (S, E, I, R, Svr, Svc, M); including four SEIR states for Ebola disease progression (susceptible S, exposed E, infectious I and refractory R) and three states for vaccination progression (two susceptible but vaccinated states $S_{VR}$ and $S_{VC}$ and an immune state M) (see Figure 1). The states Svc and Svr are assigned to individuals vaccinated by community or ring vaccination. Individuals with these states remain susceptible to Ebola exposure. After a delay of 10 days for Svr states or 28 days for Svc states, the individuals become immune by transitioning to state M. Individuals in three epidemiological states (S, Svc and Svr) are at risk of contracting Ebola from infectious individuals (I), and transitioning to the latent period (exposed state E). Transition rates from any of these three susceptible states to the exposed state depend on the contact network and increase with the number of infectious contacts. That is, if the rate of exposure per infectious contact is $t$ then the rate of exposure will be $m*t$ for a susceptible individual with $m$ infectious contacts. In our model, individuals in the exposed state E (latency) are no longer available for effective vaccination and immunization. Thus, if an individual is indicated to be vaccinated by either the ring or community vaccination programs, they may be approached by a vaccination team and given the vaccine since their exposure status is unknown, but once in the exposed class, they may not change states to the susceptible states Svc or Svr. Once exposed, nodes transition from the exposed (E) to infectious class (I) with probability $\frac{1}{\gamma}$ per day where $\gamma = 9$ days is the average incubation period [38-42] and once infectious, nodes transition from the infectious (I) to refractory class (R) with probability $\frac{1}{\lambda}$ per day, where $\lambda = 5.6$ days is the average infectious period [38-41]. Parameter definitions and values used in simulations and their sources are given in Table 1.

**Contact Network with Household and Community Structure**

As in refs. [26, 27, 37], we model the spread of Ebola on a contact network that consists of households of size H that are organized in communities of size C households. Households are arranged as a continuous line of households within an LxH grid where L is the number of households in the total contact network N. Households are indexed by their position in the line $\{h_i, h_{i+1}, ...\}$ and a network distance $\eta$ between two households $h_i$ and $h_j$ is defined as the number of households separating them ($\eta=|i-j|$). Each household has its own "community" that overlaps with the communities of other nearby households. In particular, the $i_{th}$ community is centered at the ith household and includes all of the households within a radius of the $i_{th}$ household (for a community containing C households, the community radius is Rc=(C-1)/2)). This contact network is highly mutually connected, the extent of overlap between the $i_{th}$ and $j_{th}$ communities depend on the network distance of the $i_{th}$ and $j_{th}$ households in that communities of nearby households overlap and share most households. Due to this high mutual connectivity and linear arrangement of the contact network, saturation effects build up quickly and disease progresses spatially as 1-dimensional wave through the lattice network. The size of the contact network and the size of the simulated population is unlimited since the lattice size LxH is dynamic and extends as needed as the disease and immunity extends through the population by adding households to the ends of the array.

In this version of the model, each household is designated as accessible or inaccessible as illustrated in Figure 2. This assignment is random when the household is created and once a household is labeled, this assignment does not change. During the initial construction of the contact network, and as the array is dynamically extended, households are labeled inaccessible with probability β, otherwise the household is accessible. Here we compare results as a function of the fraction of inaccessible households between 0 and 50%. (This range includes the estimate of 30% non-participating individuals that was estimated for a vaccination program when it is expanded to younger age groups in [43]; and indeed the program was approved for expansion to younger age groups in the first half of 2019 [44, 45]. This range also includes estimates of 25% for the number of cases that are missed for the ongoing DRC epidemic[46].

**Distance Dependent Transmission Rates**

At the onset of an epidemic, before the accumulation of saturation effects when all nodes of the network are susceptible except for a single infectious case, the household reproductive number $R0_H$ is the average number of secondary infections within the household of the infected individual and the community reproductive number $R0_C$ is the average number of secondary infections within the community. As in ref. [27], we set $R0_H$ =2.0 infections within the household and $R0_C$ =0.7 infections within the community, based on total $R_0$ and how they were observed to distribute within household and communities in historical Ebola epidemics [26, 38, 47, 48].

<u>Transmission rates for homogenous network transmission</u>: In the simplest case of homogeneous transmission rates, all members of a household have an equal probability of infection from an infectious contact, and all members of the broader community have an equal probability of infection from an infectious contact. Transmission rates within the household and within the community are normalized so that the reproductive number of household and community infections are always R0$_H$ and R0$_C$, respectively. Since there are H-1 household contacts within the household of a single infectious individual, and $(CH - H)$ community contacts within the community of a single infectious individual, for a fixed infectious period 1/λ, household and community transmission rates for each infectious-susceptible contact are given by:

$$t_H := \frac{R_{0H}}{\lambda\,(H-1)}, t_C := \frac{R_{0C}}{\lambda\,(C \cdot H - H)}.$$

However, as described in refs. [49, 50], for exponentially distributed infectious periods 1/λ, household and community transmission rates for each infectious-susceptible contact would be:

$$t_H := \frac{R_{0H}}{\lambda\,(H-1-R_{0H})}, t_C := \frac{R_{0C}}{\lambda\,(C \cdot H - H - R_{0C})}.$$

Since there are $(CH - H)$ community contacts, the total rate of community transmission in the network (for a single infectious individual with all other nodes susceptible) would be $(CH - H) \cdot t_C$.

Transmission rates for distance-dependent network transmission: In ref. [37], we introduced distance-dependent transmission so that within the community, the transmission rate decreases with the distance between the infectious and susceptible individuals. For any distance function f(η), transmission rates on the network may be scaled to yield a given set of reproductive numbers R0$_H$ and R0$_C$. This is accomplished by assigning values α·f(η) to each community contact and choosing α so that the sum of transmission rates (summed along every edge between an individual and every other member of the community) is equal to $(CH - H) \cdot t_C$. Initial transmission rates when all contacts are susceptible are R0$_H$ infections per household and R0$_C$ infections within the community. As the number of exposed nodes accumulate, actual transmission rates decrease due to saturation effects that depend on the radial profile of the transmission function. Thus, the spread of an infection depends on this transmission profile, as well as the household and community reproductive numbers and the network structure itself.

Here we consider relatively simple distance-dependent transmission profiles in which transmission rates decrease exponentially with distance using the function $f(\eta) = e^{-k\eta}$ that we used in ref. [37] or inversely with distance using the function $g(\eta) = \frac{1}{\eta^k}$ (the intuitive "gravity model" from transportation theory [51, 52], applied to the case of households of equal size as we have here). The normalized transmission profiles are shown in Figure 3. Other transmission functions that depend on distance may be more complex exponential forms (e.g. beta-pert distributed as in ref. [53]).

**Community and Ring Vaccination**

We model two types of vaccination. Ring vaccination is the vaccination of household members and other close contacts that are members of the community within a radius r$_{VR}$. Community vaccination is the vaccination of members of the entire community. Both types of vaccination occur on a daily basis, for flexible start and end times.

Ring vaccination: On the day D that an infectious individual within an accessible household becomes infectious, the infectious individual provides a list of their close contacts (household members and members of the community within radius r$_{VR}$). Ring vaccination occurs after delay of τ days so that vaccination occurs on D+τ, with a value that ranges from $\tau = 4$ to $\tau = 9$ days [54]. A baseline delay of $\tau = 6$ days is consistent with an estimate of 6 days (IQR: 4–9 days) for the median time-from symptom onset to effective isolation based on field data of the ongoing outbreak in DRC [54]. For reference, in prior outbreaks this delay has been shorter and estimated at 2 days based on data from the Guinea ring vaccination trial [43]. Ring vaccination begins on a number of days Vstart after a simulation begins with one infectious case. If ring vaccination has begun by day D, then members added to contact lists on day D are vaccinated on day D+τ (their state changes from S to $S_{VR}$) if they are susceptible, not yet vaccinated, and located within an accessible household. The radius $r_{VR}$ =5 households was chosen based on the household size of H=5, in agreement with the average household size in the Democratic Republic of Congo [55], to include 55 individuals. This was based on the identification of 50 contacts per infectious case during the Guinea ring vaccination trial [16]. In sensitivity analyses, we explored two additional values for the radius of ring vaccination (3 and 7 households).

Community vaccination: If community vaccination is occurring on a given day D, all community contacts of infectious individuals will be vaccinated with rate $\lambda_{vc}$ if they are susceptible, not yet vaccinated, and located within accessible households. Vaccination rates of 5% and 10% per day were studied, corresponding to a program in which on average accessible members of the community are vaccinated in 20 or 10 days. We do not model exponentially distributed immunity rates. Rather, immunization occurs for individuals with state $S_{VR}$ after 10 days and for individuals with state Svc after 28 days. It is possible that an individual is vaccinated with community vaccination and then discovers they are a close contact of an infectious individual, where ring vaccination is indicated. Because of this, a spike may occur when community vaccination is initiated, due to the comparable delay of protection for these individuals. We optimistically model that the individual can be ring vaccinated as well, either with a supplemental or additional vaccine dose, will be immunized by whichever vaccine would mature first.

This implementation of community vaccination roughly corresponds to the third ring vaccination strategy because it involves the vaccination of relatively close contacts that are in the community but outside the radius $r_{VR}$ of closest households. In our network model all community members of a household have some probability of contact. While the distance dependent transmission rates through the community would permit an objective definition of "first", "second" and "third" level contacts, we do not do that here. Instead, we partition the community into the household, the ring that receives the highest dose, and the remaining members of the community receiving the lower dose. According to a WHO graphic [25], first and second level vaccination involves the vaccination of 90 to 110 contacts with the higher dose vaccine and third level vaccination involves the vaccination of an additional 50 to 110 contacts at the lower level dose. For order of magnitude comparison, our ring vaccination involves the vaccination of 55 contacts with the higher dose and the vaccination of 350 people at the lower dose. Of note, in our network model, the community size is approximately twice as large in membership as the sum of the first, second and third rings summarized in the WHO graphic. However, a larger community size was required by our fitting of the epidemiological data (see below)).

**Simulation Description, and Fitting of the Community Size C**

For a typical simulation, the simulation is initiated on Day 1 with a single infectious individual centered within a single community. With each newly exposed individual, the contact network is dynamically extended to include the community of every exposed individual. As the simulation is run, disease and vaccination states accumulate in the network and the algorithm keeps track of the state of each node for each day. Disease and/or vaccination states progress as a linear wave through the contact network. Outbreaks die out spontaneously when all members of the contact network are susceptible, refractory or immune.

The average growth of an epidemic per day is computed as the average number of cases versus day for 500 simulations. The average number of cases for the nth day is averaged only for the number of simulations that have not yet extinguished by day n. While $R0_H$ and $R0_C$ are parameters that we consider intrinsic to the virology of Ebola, the community size C is an abstract quantity that is meant to capture the net effect of community interactions that describe the accumulation of saturation effects. In ref. [26], we found that different community sizes would result in different steady state levels for the number of cases per day, so that regions with a larger growth rate have a larger community size. The community size C must also be calibrated for each transmission profile. We choose the community size for each transmission profile by assuming that an outbreak in North Kivu would be similar in size to the 2014-16 Ebola outbreak in Liberia, if the outbreak was not curbed by vaccination. The community size and the transmission profile were also constrained by requiring that ring vaccination would be effective with 100% accessibility, since ring vaccination was highly effective in the capital [18]. Effective ring vaccination requires a steeper rather than flatter transmission profile. For such transmission profiles, a community size of C=401 was on the lower range of community sized that would have a vaccination-free incidence as high as that observed in Liberia (which had the highest sustained incidence rate of the three countries involved in the West Africa Ebola epidemic).

**Model Limitations**

We have made several simplifying assumptions. We assume that the delay to vaccination for accessible individuals is a fixed rather than variable time period that includes the time to identify the case of Ebola and any other delays to vaccination. The time period for the vaccination delay begins the day the individual transitions from exposed to infectious. We assumed that already exposed individuals could not benefit from vaccination before becoming infectious. However, there is evidence that vaccination after exposure can reduce Ebola symptoms and would likely decrease transmission of the infectious individual and for this reason in ref. [43] researchers estimated a slighter lower time to immunization of 7 rather than 10 days for the 0.5 ml ring vaccine dose. We assumed that the vaccines are 100% effective, and that they are effective after a fixed rather than random time interval. Evidence indicates that the vaccine efficacy is close to 100% [16]. By assuming a 100% vaccine effectiveness, we assume that the effects of non-participation in vaccination or unvaccinated community contacts are a much larger effect than a small percentage of vaccine failures.

There are many unknowns regarding the parametrization of the household community structure. We chose a community size that would match the incidence observed in Liberia during the West African Ebola epidemic, in the absence of vaccination, but this community size depends on the transmission profile as a function of distance that is unknown. The community size and the transmission profile together should be interpreted as a summary of complex network interactions that result in an observed incidence and are not constrained by the currently available data.

One of the most significant limitations of our model is that we model a random distribution of inaccessible households (also as in ref. [43]), whereas they are likely to be clustered within communities. A simplifying assumption we have made is that there are only two types of households, "accessible" and "inaccessible", so that we assume that individuals that do not provide contact lists, for whatever reason, are also individuals that do not participate in vaccination, for whatever reason. However, if we separated such groups, we would have very little data to decide how these groups would overlap. We assume that accessibility status is assigned at the household level, rather than at the individual or community level, and is the same for either community vaccination or ring vaccination. Whereas certainly individuals in a household may hold different attitudes and different levels of geographic accessibility, and individual behaviors may be different depending on whether contact tracing and vaccination teams are visiting in the household or located in the community, it seems a reasonable first approximation that all the outcomes will cluster at the household levels. As mentioned above, there is likely clustering at the community level as well, and in principle, there should be non-random assignment of accessibility for households within communities. Vaccination data from the MOH situation reports have suggested higher rates of vaccination in certain health zones, which suggests heterogeneous inaccessibility [56]. Likewise, violence targeting healthcare workers has been concentrated in certain zones. However, we again do not have the data to specify this clustering and have committed to a random distribution. An obviating consideration is that our modeling results should be interpreted as predictions applied to a relatively small geographic region, for subpopulations of approximately 5000 households over 18 months. If clustering occurs at larger scales (e.g., such as at the health zone level) our results apply to subpopulations that might have a relatively low or relatively high accessibility, but homogenously, within that subpopulation.

Distance-dependent transmission rates are meant to account for the net effect of all routes of transmission, including, for example, nosocomial and funeral transmission. We have not modeled these transmission rates (nor the network compartments) separately, which would be required for modeling the effect of time-dependent improvements of common interventions on transmission rates.

Finally, we simulate epidemics in the context of community transmission in the absence of any long-range links, so we cannot capture the effects of missed contacts that seed outbreaks in new locations (and the initial faster growth of those outbreaks in naïve communities). In our simplified network model, households do not necessarily map to family residences, but instead households and highly distance-dependent transmission rates between nearby households should include a variety of close contact structures such as multi-generational households, health centers, school and work networks. The linearity of the household-community structures provides a conveniently simplified network model but is highly unrealistic in terms of real human contact networks.

**Table 1. Parameter values used in simulations.** Description of each model parameter, the value or range that is used, and the reference source for the value that is used if applicable.

| Parameter | Description | Parameter Value (Range) | Source |
|---|---|---|---|
| H | Household Size | 5 | [26, 27] and [37] |
| C | Community Size | 401 | – |
| B | Fraction of households that are inaccessible. | 0.0-0.6 | – |
| $R_{0H}$ | Household reproductive number | 2.0 | [27] |
| $R_{0C}$ | Community reproductive number | 0.7 | [27] |
| $1/\gamma$ | Average incubation period | 9 days | [42] |
| $1/\lambda$ | Average infectious period | 5.6 days | [39, 47] |
| $r_{VR}$ | Radius of ring vaccination (number of households) | 5 (3, 7) | [16] |
| $\lambda_{VC}$ | Rate of community vaccination | 0.05, 0.1 | – |
| $\tau$ | Days to vaccinate an individual with the ring vaccination program | 6 (4, 9) | [43, 54] |
| $d_{VR}$ | Days till immunity for 0.5 ml vaccine | 10 | [20, 22] |
| $d_{VC}$ | Days till immunity for 0.2 ml vaccine | 28 | [20, 22] |
| $Vstart_r$ | Start of the ring vaccination campaign | 0-9 months | |
| $Vstart_c$ | Start of the community vaccination campaign | 0-9 months | |

**Results**

Incidence and extinction with no vaccine interventions

In the absence of vaccination, our calibrated model predicts that case incidence grows during the first 10 serial intervals and then surviving epidemic realizations reach a stable case incidence around 40-43 cases per day (Figure 4). A significant fraction of epidemic realizations dies out; about 22-30% of the epidemic realizations spontaneously extinguish early on, within the first 30 days of the simulation (Figure 4). Once epidemic realizations have persisted several serial intervals, they become robust to extinction with a very low extinction rate.

Incidence and extinction with ring vaccination

Our results indicate that the impact of a ring vaccination strategy depends on the fraction of inaccessible households. Using our baseline parameter values (Table 1), in the best-case scenario when all households are accessible ($\beta=0$), the probability of epidemic control increases over time with a ring vaccination strategy that starts 30 days after epidemic onset (Figure 5). Since epidemic realizations are likely to end spontaneously within the first 30 days, the effect of ring vaccination applied after 30 days shows the effect of ring vaccination on the fraction of remaining epidemics that would otherwise likely persist. Data from the curves in Figure 5 showing this increase in the probability of epidemic extinction over time are summarized in Figure 6. Over 12 and 18 months, the supplemental probability of epidemic containment with no further cases of Ebola with ring vaccination for a fully accessible population is 22-32% and 33-44%, respectively. At a 10% level of household inaccessibility, the probability of epidemic control decreases relative to the case of full accessibility (curves in Figure 5, Panel A) but the probability that the epidemic is contained with no further cases of Ebola still slowly accumulates over time. At higher (>10%) levels of household inaccessibility, the pattern is qualitatively different. The probability of epidemic control no longer increases with the application of ring vaccination and rapidly saturates around 19-30%, just as it would without the application of ring vaccination. However, the trajectory of outbreak incidence conditional on non-extinction stabilizes around 8-22 cases per day, which is markedly lower than the case of no ring vaccination (Figure 5).

These results for vaccination applied at 30 days show that for even low fractions of inaccessible households and a significant delay to vaccinating individuals ($\tau = 6$ days), ring vaccination has little effect on ending an epidemic but substantially decreases the size of the epidemic. This pattern is robust to the timing of start of the ring vaccination program: when ring vaccination is applied to an established epidemic wave after 9 months, there is no effect on the lifetime of the epidemic if the fraction of accessible households is ≥20%, but the steady state incidence decreases (Figure 7). Ring vaccination applied after 9 months decreases steady state incidence to the same values as ring vaccination applied earlier, after 30 days (Figure 5).

Results from sensitivity analyses indicate that increasing the radius of ring vaccination from 5 households to 7 significantly enhances the probability of epidemic control when levels of household inaccessibility are below 20% (Figure 6). However, decreasing the delay to vaccinating contacts from 6 to 4 days only increases the probability of epidemic control by about 10% for the best case scenario when all households are accessible (Figure 6), but it does not influence the probability of epidemic control even for low levels of household inaccessibility (as low as 10%).

<u>Community vaccination that supplements a ring vaccination strategy can significantly increase the probability of epidemic control.</u>

For the scenario resembling the DRC Ebola epidemic where a ring vaccination strategy starts one week after epidemic onset and is subsequently supplemented with community vaccination 9 months later, we found that the community vaccination strategy substantially increases the probability of extinction, after a delay, and can substantially decrease the endemic state (Figure 8). However, the probability of achieving epidemic containment again appears to not be improved when the fraction of inaccessible household is too high (e.g., 50% inaccessible; Figure 8). Figure 9 illustrates the effects of the community vaccination rates on case incidence and the probability of epidemic control. The supplemental effect of community vaccination on the probability of epidemic extinction for populations with different levels of inaccessible households ranging from 0 to 50% are summarized in Figure 10. After 1.5 years of disease transmission, our model-based results indicate that the probability of epidemic control is at 60-71% for a 30% level of household inaccessibility and 30-41% for a 40% household inaccessibility level based on a community vaccination rate of 10% per day.

As shown in Figures 9 and 10, supplemental community vaccination has a substantial effect on decreasing the daily incidence of cases and increasing the rate of extinction. While community vaccination has a near-immediate effect on decreasing case incidence, the effect of community vaccination on the rate of extinction occurs with a delay, from 3 months to as long as a year, that varies with the fraction of accessible households and the daily rate of immunization (Figures 9 and 10). For the longest delay of one year for the case of 40% inaccessible households, the increase in the rate of extinction is striking for showing an abrupt rise after such a long delay. For all conditions evaluated in Figures 9 and 10, the rate of extinction becomes non-negligible and the probability of extinction begins increasing only when the case incidence drops to a small number of cases (2-4 cases). This provides an explanation for the year-long delay observed for the case of 40% inaccessible households : the daily incidence drops at a slower rate so that it takes a year for the incidence to fall to a sufficiently low level of cases.  For populations where the fraction of inaccessible households is higher and the probability of containment does not increase at all; the daily incidence decreases but never decreases to such low levels before reaching a steady state number of cases.

**Discussion**

In this paper we have employed an individual-level stochastic transmission model to evaluate ring and community vaccination strategies for containing Ebola epidemics in the context of varying levels of community accessibility to public health interventions. Our individual-based model, which was calibrated based on the transmission dynamics of the 2014-16 Western African Ebola epidemic in the absence of vaccination, incorporates four sources of heterogeneity including a proportion of the population that is inaccessible for effective contact tracing and vaccination efforts, two levels of population mixing resembling household and community transmission, two types of vaccine doses with different time periods until immunity, and spatial dependence on transmission rates. Our findings indicate that ring vaccination, which targets a radius of contacts for each infectious individual, is an effective intervention to contain Ebola epidemics at low levels of household inaccessibility (<10%) in the presence of significant delays to vaccinating contacts. At higher levels of household inaccessibility, ring vaccination could be useful for reducing the disease endemicity level, but it is no longer an effective intervention to ensure outbreak containment in the absence of any other interventions. This is in agreement with other modeling results that ring vaccination may not be sufficient for containing outbreaks with higher reproductive numbers (15), in the range we have here, and when a fraction of individuals in the transmission chain are missed and do not participate in ring vaccination (39).

In this study, our model was designed to evaluate the impact of an intermediate-scale "3$^{rd}$ ring" level of community vaccination and the effect of variable delays to vaccination and variable levels of community participation, especially in the context of two vaccine doses that would require different times to achieve immunity. Our household-community network model is especially suited for an intermediate scale level of vaccination since communities are a network unit in our model centered at each household. For adjacent households, communities modulate by the inclusion and exclusion of one household at a time as the communities linearly arrange through the network. Note that this network definition of minutely modulating, overlapping communities, while natural for defining a 3rd level of contacts, is different from that of the usual notion of communities that divide a population into relatively disjoint groups, with some fraction of network connections between them.

We found that community vaccination strategies that supplement a ring vaccination strategy can speed up and enhance the probability of epidemic containment. The substantial impact of the community vaccination (figure 8-10) shows that the longer interval to immunity of the lower vaccine dose does not harm containment efforts; that is, in our simulations the vaccination wave moves ahead of the outbreak wave even with the longer time period to immunity involved. There are two careful caveats with this. First, community vaccination with the lower vaccine dose has a substantial impact compared to no dose (that is, no community vaccination). We modeled community vaccination for the case of a 0.2 ml vaccine dose.

Second, a small number of individuals who have been tagged for community vaccination and have taken the lower dose vaccine will discover within 18 days (the difference between 28 days and 10 days) that they are within the ring of an infected individual. We modeled that the individual would take the higher dose vaccine (or a booster) vaccine since it would be important to provide the faster protection. In simulations where we did not do this, there was an initial spike in cases the first couple weeks that community vaccination was implemented, until community vaccination had time to reduce the incidence.

While community vaccination decreases transmission and increases the probability of the outbreak extinguishing, reliable containment of the epidemic still occurs only for moderate levels of household inaccessibility (e.g., <50%). For instance, our results for the scenario motivated by the DRC Ebola epidemic predict a low but steady probability of epidemic containment of about 5% per month for a 30% household inaccessibility level (Figure 9). More generally, our results highlight the critical need to enhance community engagement to public health interventions while offering a safe and secure environment to the population in order to increase the effectiveness of control interventions to ensure outbreak containment.

Active local conflict and lack of community trust of the government and public health authorities hamper containment of epidemics driven by person-to-person transmission [9]. In this context, efforts to increase community engagement could enhance the effectiveness of contact tracing activities and vaccination acceptance rates. [7]. Findings from a recent survey in DRC indicate that low institutional trust is associated with a decreased likelihood of adopting preventative behaviors, including seeking formal health services at hospitals or health centres [9]. Importantly, there is some evidence that Ebola outbreak in DRC is being perceived in the region as a political tool against certain groups while some other groups benefit politically and financially from the ongoing epidemic [57]. This underscores the need to ensure an objective Ebola response that is entirely isolated from political biases. Indeed, prior work suggests that the implementation of integrative communal approaches can help improve epidemiological surveillance and enhance the adoption of Ebola preventive measures [7, 59-61].

In contrast to the simple SIR compartmental transmission models based on homogenous mixing assumptions that support bell-shaped epidemic trajectories [62, 63] [64, 65], our spatially structured stochastic model employed in this study has been able to successfully capture stationary disease waves where the virus moves through the host population over time. This transmission pattern is qualitatively similar to those of the ongoing Ebola epidemic in DRC [66, 67]. In particular, they achieve relatively steady incidence levels as Ebola spreads from one community to another, and the total incidence can be interpreted as the super-positioning of the cases from a discrete number of communities involved rather than explosive exponential growth. However, our model does not predict sporadic increases in case incidence, as in the most recent surge in case incidence [68], which is not surprising as our model does not incorporate time-dependent changes in the effectiveness of control interventions tied to violent attacks to healthcare workers and public health infrastructure. From an epidemic modeling perspective, our results underscore the need to capture an appropriate spatial structure in models of disease transmission [26, 31-33, 69, 70].

Such considerations may be more important for infectious diseases that are transmitted via close contact such as Ebola and HIV.

In summary, we found that the role of vaccination strategies in containing Ebola epidemics significantly depends on the level of community inaccessibility using an individual-level stochastic transmission model that successfully captures stationary disease waves that are qualitatively similar to those of the ongoing Ebola epidemic in DRC. For lower levels of inaccessibility, the probability of containment increases over time. For higher levels, vaccination strategies investigated in this study are not expected to contain the epidemic, but they help reduce incidence levels, which saves lives, makes the epidemic easier to contain and reduces spread to other communities. This qualitative change occurs for both types of vaccination strategies: ring vaccination is effective for containing an outbreak until the level of inaccessibility exceeds approximately 10%, a combined ring and community vaccination strategy is effective until the level of inaccessibility exceeds approximately 50%. In order to enhance the effectiveness of control interventions, it is crucial to ensure community engagement to public health interventions.

**Conflict of interest**

The authors declare no conflicts of interest.

**Funding**
NSF grant 1414374 as part of the joint NSF-NIH-USDA Ecology and Evolution of Infectious Diseases program (GC).

**References**

1. Cohen J. Ebola outbreak continues despite powerful vaccine. Available from: https://science.sciencemag.org/content/364/6437/223. Science. 2019.
2. Cohen J. Ebola outbreak continues despite powerful vaccine. 2019:223-.
3. 2014 Ebola Outbreak in West Africa - Reported Cases Graphs 2016 [cited 2017 March 23]. Available from: https://www.cdc.gov/vhf/ebola/outbreaks/2014-west-africa/cumulative-cases-graphs.html.
4. Pandey A, Atkins KE, Medlock J, Wenzel N, Townsend JP, Childs JE, et al. Strategies for containing Ebola in west Africa. Science. 2014;346(6212):991-5.
5. Olu OO, Lamunu M, Nanyunja M, Dafae F, Samba T, Sempiira N, et al. contact Tracing during an Outbreak of ebola Virus Disease in the Western area Districts of sierra leone: lessons for Future ebola Outbreak response. Frontiers in Public Health. 2016;4.
6. Martín AC, Derrough T, Honomou P, Kolie N, Diallo B, Koné M, et al. Social and cultural factors behind community resistance during an Ebola outbreak in a village of the Guinean Forest region, February 2015: a field experience. International health. 2016:ihw018.
7. Marais F, Minkler M, Gibson N, Mwau B, Mehtar S, Ogunsola F, et al. A community-engaged infection prevention and control approach to Ebola. Health Promot Int. 2016;31(2):440-9. Epub 2015/02/15. doi: 10.1093/heapro/dav003. PubMed PMID: 25680362.
8. Nyenswah TG, Kateh F, Bawo L, Massaquoi M, Gbanyan M, Fallah M, et al. Ebola and Its Control in Liberia, 2014-2015. Emerging infectious diseases. 2016;22(2):169-77. Epub 2016/01/27. doi: 10.3201/eid2202.151456. PubMed PMID: 26811980; PubMed Central PMCID: PMCPMC4734504.
9. Vinck P, Pham PN, Bindu KK, Bedford J, Nilles EJ. Institutional trust and misinformation in the response to the 2018-19 Ebola outbreak in North Kivu, DR Congo: a population-based survey. The Lancet Infectious diseases. 2019;19(5):529-36. Epub 2019/04/01. doi: 10.1016/S1473-3099(19)30063-5. PubMed PMID: 30928435.
10. Nguyen VK. An Epidemic of Suspicion - Ebola and Violence in the DRC. The New England journal of medicine. 2019;380(14):1298-9. Epub 2019/03/07. doi: 10.1056/NEJMp1902682. PubMed PMID: 30840790.

11. Claude KM, Underschultz J, Hawkes MT. Ebola virus epidemic in war-torn eastern DR Congo. Lancet. 2018;392(10156):1399-401. Epub 2018/10/10. doi: 10.1016/S0140-6736(18)32419-X. PubMed PMID: 30297137.

12. Vera Y, Yeung J. Ebola: DR Congo confirms first case in city of Goma on border with Rwanda. Available from: https://www.cnn.com/2019/07/14/health/ebola-outbreak-goma-drc/index.html (accessed on 15 July 2019). 2019.

13. WHO, Ebola Virus Disease, Democratic Republic of the Congo, External Situation Report 45. 2019: World Health Organization. Available from: https://apps.who.int/iris/bitstream/handle/10665/325242/SITREP_EVD_DRC_UGA_20190612-eng.pdf?ua=1 (accessed on 12 June 2019).

14. WHO. Ebola virus disease – Democratic Republic of the Congo
Disease outbreak news. 2019  June 14, 2019]; Available from: https://www.who.int/csr/don/13-june-2019-ebola-drc/en/. (accessed on 15 June 2019).

15. Kivu Security Tracker. Available from: https://kivusecurity.org/map#.

16. Henao-Restrepo AM, Longini IM, Egger M, Dean NE, Edmunds WJ, Camacho A, et al. Efficacy and effectiveness of an rVSV-vectored vaccine expressing Ebola surface glycoprotein: interim results from the Guinea ring vaccination cluster-randomised trial. Lancet. 2015;386(9996):857-66. doi: 10.1016/S0140-6736(15)61117-5. PubMed PMID: 26248676.

17. Merler S, Ajelli M, Fumanelli L, Parlamento S, Pastore YPA, Dean NE, et al. Containing Ebola at the Source with Ring Vaccination. PLoS Negl Trop Dis. 2016;10(11):e0005093. Epub 2016/11/03. doi: 10.1371/journal.pntd.0005093. PubMed PMID: 27806049; PubMed Central PMCID: PMCPMC5091901.

18. Wells CR, Pandey A, Parpia AS, Fitzpatrick MC, Meyers LA, Singer BH, et al. Ebola vaccination in the Democratic Republic of the Congo. Proceedings of the National Academy of Sciences of the United States of America. 2019;116(20):10178-83. Epub 2019/05/01. doi: 10.1073/pnas.1817329116. PubMed PMID: 31036657.

19. WHO. Ebola virus disease – Democratic Republic of the Congo. 2018  [cited 2018 August 19]; Available from: https://www.who.int/csr/don/17-august-2018-ebola-drc/en/.

20.  The report of the SAGE April 2019 will be published in the WHO Weekly Epidemiological Record (www.who.int/wer/en/) on 30 May 2019.

21. Maxmen, A. Ebola detectives race to identify hidden sources of infection as outbreak spreads. Available from: https://www.nature.com/articles/d41586-018-07618-0 (accessed on 23 May 2019).

22. Cohen J. DRC expands Ebola vaccine campaign as cases mount rapidly. Science.
23. World Health Organization. Interim Recommendations on Vaccination against Ebola Virus Disease (EVD). Available from: https://www.who.int/immunization/policy/position_papers/interim_ebola_recommendations_may_2019.pdf?ua=1 (accessed on 20 May 2019).
24. Cuddy A. Ebola: How a disease is prevented from spreading. The BBC News. Available from: https://www.bbc.com/news/world-africa-49109478 (accessed on 28 July 2019).
25. Branswell H. WHO broadens the pool of people who can get the Ebola vaccine. STAT. Available from: https://www.statnews.com/2019/05/07/who-broadens-eligibility-ebola-vaccine/ (accessed on 7 May 2019). 2019.
26. Kiskowski M. Three-Scale Network Model for the Early Growth Dynamics of 2014 West Africa Ebola Epidemic. . PLOS Currents Outbreaks. 2014;doi: 10.1371/currents.outbreaks.b4690859d91684da963dc40e00f3da81. doi: doi: 10.1371/currents.outbreaks.c6efe8274dc55274f05cbcb62bbe6070.
27. Kiskowski M, Chowell G. Modeling household and community transmission of Ebola virus disease: epidemic growth, spatial dynamics and insights for epidemic control. Virulence. 2015;7(2):63-73. doi: doi: 10.1080/21505594.2015.1076613. PubMed PMID: 26399855.
28. BBC News. Half of Ebola cases in DR Congo 'unidentified'. Available from: https://www.bbc.com/news/world-africa-49212116 (accessed on 12 August 2019).
29. Tildesley MJ, Savill NJ, Shaw DJ, Deardon R, Brooks SP, Woolhouse ME, et al. Optimal reactive vaccination strategies for a foot-and-mouth outbreak in the UK. Nature. 2006;440(7080):83-6. doi: 10.1038/nature04324. PubMed PMID: 16511494.
30. Ferguson NM, Donnelly CA, Anderson RM. The foot-and-mouth epidemic in Great Britain: pattern of spread and impact of interventions. Science. 2001;292(5519):1155-60. Epub 2001/04/17. doi: 10.1126/science.1061020 1061020 [pii]. PubMed PMID: 11303090.
31. Browne C, Gulbudak H, Webb G. Modeling contact tracing in outbreaks with application to Ebola. Journal of theoretical biology. 2015;384:33-49. doi: 10.1016/j.jtbi.2015.08.004. PubMed PMID: 26297316.
32. Lau MSY, Gibson GJ, Adrakey H, McClelland A, Riley S, Zelner J, et al. A mechanistic spatio-temporal framework for modelling individual-to-individual transmission-With an application to the 2014-2015 West Africa Ebola outbreak. PLoS computational biology. 2017;13(10):e1005798. Epub 2017/10/31. doi:

10.1371/journal.pcbi.1005798. PubMed PMID: 29084216; PubMed Central PMCID: PMCPMC5679647.

33. Yang W, Zhang W, Kargbo D, Yang R, Chen Y, Chen Z, et al. Transmission network of the 2014-2015 Ebola epidemic in Sierra Leone. Journal of the Royal Society, Interface / the Royal Society. 2015;12(112). Epub 2015/11/13. doi: 10.1098/rsif.2015.0536. PubMed PMID: 26559683; PubMed Central PMCID: PMCPMC4685836.

34. Greenhalgh D. Optimal control of an epidemic by ring vaccination. Communications in Statistics Stochastic Models. 1986;2(3):339--63. doi: doi: 10.1080/15326348608807041.

35. Muller J, Schonfisch B, Kirkilionis M. Ring vaccination. Journal of mathematical biology. 2000;41(2):143-71. PubMed PMID: 11039695.

36. Kretzschmar M, van den Hof S, Wallinga J, van Wijngaarden J. Ring vaccination and smallpox control. Emerging infectious diseases. 2004;10(5):832-41. Epub 2004/06/18. PubMed PMID: 15200816.

37. Chowell G, Kiskowski M. Modeling ring-vaccination strategies to control Ebola virus disease epidemics. In: Chowell G, Hyman JM, editors. Mathematical modelling for emerging and reemerging infectious diseases: Springer; 2016.

38. Chowell G, Hengartner NW, Castillo-Chavez C, Fenimore PW, Hyman JM. The basic reproductive number of Ebola and the effects of public health measures: the cases of Congo and Uganda. Journal of theoretical biology. 2004;229(1):119-26. Epub 2004/06/05. doi: 10.1016/j.jtbi.2004.03.006 S0022519304001092 [pii]. PubMed PMID: 15178190.

39. Legrand J, Grais RF, Boelle PY, Valleron AJ, Flahault A. Understanding the dynamics of Ebola epidemics. Epidemiology and infection. 2007;135(4):610-21. doi: 10.1017/S0950268806007217. PubMed PMID: 16999875; PubMed Central PMCID: PMC2870608.

40. Lekone PE, Finkenstadt BF. Statistical inference in a stochastic epidemic SEIR model with control intervention: Ebola as a case study. Biometrics. 2006;62(4):1170-7. doi: 10.1111/j.1541-0420.2006.00609.x. PubMed PMID: 17156292.

41. Eichner M, Dowell SF, Firese N. Incubation period of ebola hemorrhagic virus subtype zaire. Osong Public Health Res Perspect. 2011;2(1):3-7. doi: 10.1016/j.phrp.2011.04.001. PubMed PMID: 24159443; PubMed Central PMCID: PMC3766904.

42. Team WHOER. Ebola Virus Disease in West Africa - The First 9 Months of the Epidemic and Forward Projections. The New England journal of medicine. 2014;371(16):1481-95. doi: 10.1056/NEJMoa1411100. PubMed PMID: 25244186.

43. Kucharski AJ, Eggo RM, Watson CH, Camacho A, Funk S, Edmunds WJ. Effectiveness of Ring Vaccination as Control Strategy for Ebola Virus Disease. Emerging infectious diseases. 2016;22(1):105-8. Epub 2015/12/23. doi: 10.3201/eid2201.151410. PubMed PMID: 26691346; PubMed Central PMCID: PMCPMC4696719.

44. ReliefWeb Report. UNICEF DR Congo Ebola Situation Report North Kivu and Ituri - 03 February 2019. Available from: https://reliefweb.int/report/democratic-republic-congo/unicef-dr-congo-ebola-situation-report-north-kivu-and-ituri-03.

45. Ilunga Kalenga O, Moeti M, Sparrow A, Nguyen VK, Lucey D, Ghebreyesus TA. The Ongoing Ebola Epidemic in the Democratic Republic of Congo, 2018-2019. The New England journal of medicine. 2019. Epub 2019/05/30. doi: 10.1056/NEJMsr1904253. PubMed PMID: 31141654.

46. Branswell, H. WHO sees progress in Ebola response, but others see a grimmer reality. StatNews. Available from: https://www.statnews.com/2019/06/06/who-sees-progress-in-ebola-response-but-others-see-a-grimmer-reality/ (accessed on 06/06/2019). 2019.

47. Chowell G, Nishiura H. Transmission dynamics and control of Ebola virus disease (EVD): a review. BMC medicine. 2014;12(1):196. doi: 10.1186/s12916-014-0196-0. PubMed PMID: 25300956.

48. Althaus CL. Estimating the reproduction number of Zaire ebolavirus (EBOV) during the 2014 outbreak in West Africa. PLOS Currents Outbreaks Edition 1 doi: 101371/currentsoutbreaks91afb5e0f279e7f29e7056095255b288. 2014.

49. Lloyd AL VS, Cintrón-Arias A. Infection dynamics on small-world networks. Contemp Math 2006;410:209.

50. Keeling MJ, Grenfell BT. Individual-based perspectives on R(0). Journal of theoretical biology. 2000;203(1):51-61. Epub 2000/03/30. doi: 10.1006/jtbi.1999.1064. PubMed PMID: 10677276.

51. Xia YC, Bjørnstad ON, Grenfell BT. Measles metapopulation dynamics: a gravity model for epidemiological coupling and dynamics. Am Nat 2004;164:267-81.

52. Viboud C, Bjornstad ON, Smith DL, Simonsen L, Miller MA, Grenfell BT. Synchrony, waves, and spatial hierarchies in the spread of influenza. Science. 2006;312(5772):447-51. Epub 2006/04/01. doi: 1125237 [pii] 10.1126/science.1125237. PubMed PMID: 16574822.

53. Durr S, Ward MP. Development of a Novel Rabies Simulation Model for Application in a Non-endemic Environment. PLoS Negl Trop Dis. 2015;9(6):e0003876. Epub 2015/06/27. doi: 10.1371/journal.pntd.0003876. PubMed PMID: 26114762; PubMed Central PMCID: PMCPMC4482682.

54. WHO Outbreak News: 06/06/2019. Ebola virus disease – Democratic Republic of the Congo. Available from: https://www.who.int/csr/don/06-june-2019-ebola-drc/en/ (accessed on 06/15/2019).

55. Ministère du Plan et Suivi de la Mise en œuvre de la Révolution de la Modernité (MPSMRM), Ministère de la Santé
Publique (MSP) and ICF International. 2014. Democratic Republic of Congo Demographic and Health Survey 2013-14:
Key Findings. Rockville, Maryland, USA: MPSMRM, MSP et ICF International. Available from: https://dhsprogram.com/pubs/pdf/SR218/SR218.e.pdf. 2014.

56. Soucheray S. WHO: Ebola spread in DRC still 'moderate'. CIDRAP. Available from: http://www.cidrap.umn.edu/news-perspective/2019/03/who-ebola-spread-drc-still-moderate (accessed on 10 April 2019). 2019.

57. Trapido J. Ebola: public trust, intermediaries, and rumour in the DR Congo. The Lancet Infectious diseases. 2019. Epub 2019/04/01. doi: 10.1016/S1473-3099(19)30044-1. PubMed PMID: 30928434.

58. Aizenman N. Why The Ebola Outbreak In The Democratic Republic Of Congo Keeps Getting Worse. NPR News. Available from: https://www.npr.org/2019/05/23/725789837/why-the-ebola-outbreak-in-the-democratic-republic-of-congo-keeps-getting-worse (accessed on 23 May 2019).

59. Tambo E, Ugwu EC, Ngogang JY. Need of surveillance response systems to combat Ebola outbreaks and other emerging infectious diseases in African countries. Infect Dis Poverty. 2014;3:29. Epub 2014/08/15. doi: 10.1186/2049-9957-3-29. PubMed PMID: 25120913; PubMed Central PMCID: PMCPMC4130433.

60. Thiam S, Delamou A, Camara S, Carter J, Lama EK, Ndiaye B, et al. Challenges in controlling the Ebola outbreak in two prefectures in Guinea: why did communities continue to resist? Pan Afr Med J. 2015;22 Suppl 1:22. Epub 2016/01/08. doi: 10.11694/pamj.supp.2015.22.1.6626. PubMed PMID: 26740850; PubMed Central PMCID: PMCPMC4695515.

61. Buseh AG, Stevens PE, Bromberg M, Kelber ST. The Ebola epidemic in West Africa: challenges, opportunities, and policy priority areas. Nurs Outlook. 2015;63(1):30-40. Epub 2015/02/04. doi: 10.1016/j.outlook.2014.12.013. PubMed PMID: 25645480.

62. Anderson RM, May RM. Infectious diseases of humans. Oxford: Oxford University Press; 1991.

63. Hethcote HW. The mathematics of infectious diseases. SIAM review. 2000;42(4):599-653.

64. Brauer F. Some simple epidemic models. Mathematical biosciences and engineering : MBE. 2006;3(1):1-15. PubMed PMID: 20361804.

65. Kermack WO, McKendrick AG. Contributions to the mathematical theory of epidemics: IV. Analysis of experimental epidemics of the virus disease mouse ectromelia. J Hyg (Lond). 1937;37(2):172-87. PubMed PMID: 20475371; PubMed Central PMCID: PMC2199429.

66. Ebola virus disease – Democratic Republic of the Congo. Disease outbreak news: Update for 23 April 2019. Available from: https://www.who.int/csr/don/23-may-2019-ebola-drc/en/ (accessed on 24 May 2019).

67. Tariq A, Roosa K, Mizumoto K, Chowell G. Assessing reporting delays and the effective reproduction number: The Ebola epidemic in DRC, May 2018-January 2019. Epidemics. 2019;26:128-33. PubMed PMID: WOS:000463187400013.

68. WHO. (2019). Ebola Virus Disease, Democratic Republic of Congo, External Situation Reports. April 24, 2019.   Retrieved April 24, 2019. Available from:https://apps.who.int/iris/bitstream/handle/10665/312080/SITREP-EVD-DRC-20192404-eng.pdf?ua=1.

69. Sattenspiel L, Dietz K. A structured epidemic model incorporating geographic mobility among regions. Mathematical biosciences. 1995;128(1-2):71-91. PubMed PMID: 7606146.

70. Merler S, Ajelli M, Fumanelli L, Gomes MF, Piontti AP, Rossi L, et al. Spatiotemporal spread of the 2014 outbreak of Ebola virus disease in Liberia and the effectiveness of non-pharmaceutical interventions: a computational modelling analysis. The Lancet Infectious diseases. 2015;15(2):204-11. doi: 10.1016/S1473-3099(14)71074-6. PubMed PMID: 25575618.

**Figures**

**Figure 1.** The seven epidemiological states for individuals in the population and the transitions between states. Susceptible (S) and individuals vaccinated with either the ring or community dose of vaccine (Svr, SVc) may become exposed (E). Once exposed, individuals eventually become infectious and then refractory. Vaccinated individuals that do not become exposed eventually become immunized (M).

| | |
|---|---|
| S | Susceptible to Ebola |
| $Sv_c$ | Susceptible to Ebola but vaccinated with community dose of the vaccine |
| $Sv_r$ | Susceptible to Ebola but vaccinated with ring dose of the vaccine |
| E | Exposed to Ebola |
| I | Infectious |
| R | Refractory after an Ebola infection |
| M | Immunized by the vaccine and no longer susceptible to Ebola |

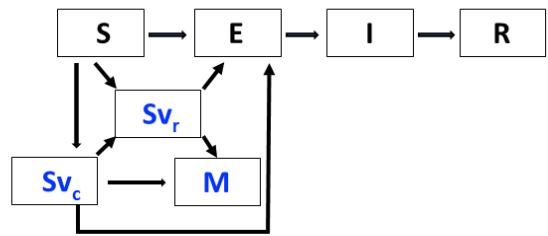

**Figure 2.** Description of the contact network. A) The contact network is initialized with a single infectious individual (red node) and that individual's community. For this illustrative schematic, the household size H=5 and the community size is small with C=25. The accessibility (white nodes) or inaccessibility (gray nodes) of a household is assigned randomly during the construction of the contact network. The index case happens to be in an inaccessible household: the first case of the simulation does not supply a contact list, and no contacts are vaccinated. B) After several timesteps, the infectious index case exposes another member of their household (blue node in the $h_{ith}$ household) and a member of their community (blue node at a distance of 8 households). When the community member is exposed in the $h_{i+8th}$ household, the total contact network is extended by 8 households to include their community (the $h_{i+8th}$ community is outlined with a black rectangle). C) The $h_{i+8th}$ household is accessible, so once the exposed individual becomes infectious a contact list is provided (green nodes) of households within a radius Rv=2. One household within this radius is excluded since it is inaccessible. The members of this household are not included on contact lists, do not participate in vaccination, and/or are not geographically accessible to the vaccination teams, etc.

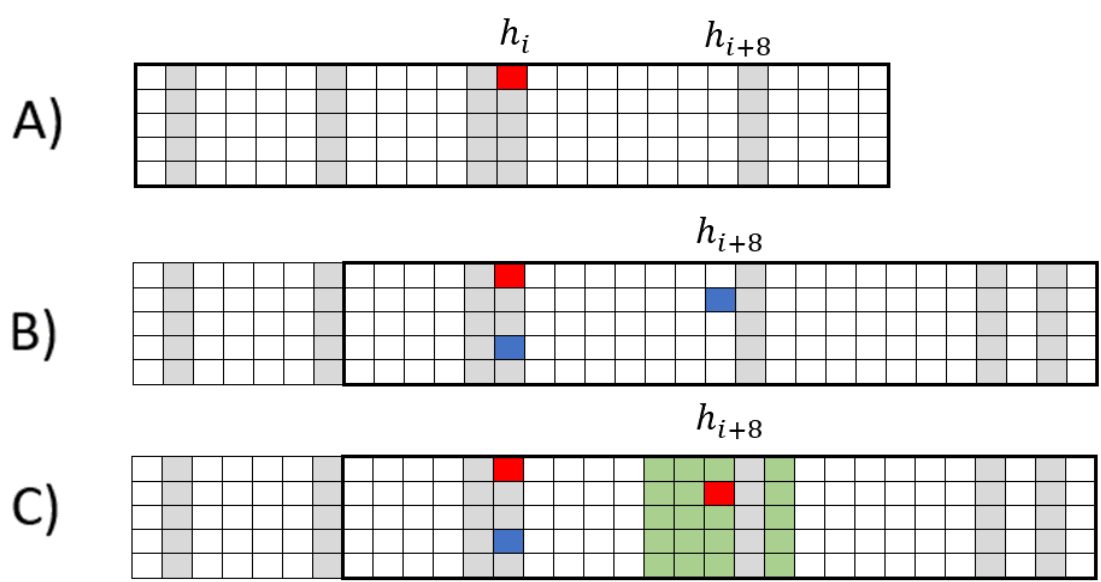

**Figure 3**: Community transmission profiles and steady states in the absence of vaccination. A) Transmission rates as a function of distance from an infectious individual for an inverse transmission profile (left) and exponential transmission profile (right) with community reproductive number R$_{0C}$=0.7 for both profiles. The inverse distance function is $g(\eta) = \frac{1}{\eta^{0.93}}$ and the exponential distance function is $f(\eta) = e^{-0.02\eta}$ for the community size C=401. B). For this community size and transmission profile parameters, a steady state incidence comparable to that of the 2014-16 Ebola epidemic in Liberia is achieved when simulating epidemics in the absence of vaccination (see Figure 4).

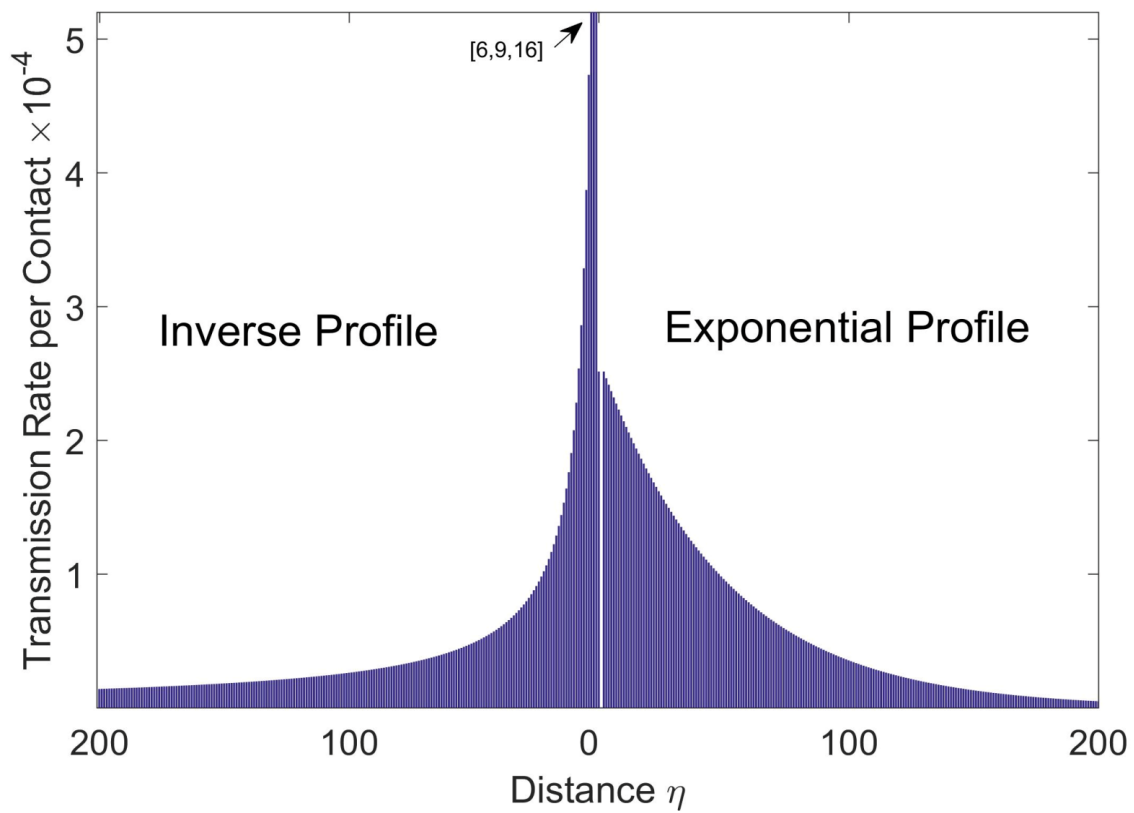

**Figure 4.** Probability of epidemic extinction and the baseline mean daily incidence curves in the absence of interventions for the exponential and gravity transmission profiles that define transmission rates as a function of the distance from an infectious individual (see also Figure 3). For each transmission profile, the model was calibrated by choosing the community size for each transmission profile by assuming that the ongoing Ebola outbreak in North Kivu would be similar in size to the 2014-16 Ebola outbreak in Liberia in the absence of vaccination. Outbreaks propagate through the population as spatial waves with an endemic state. The average number of cases for the $n_{th}$ day is averaged only for the number of simulations that have not yet extinguished by day n. The error bars show the 95% confidence intervals for the mean daily incidence.

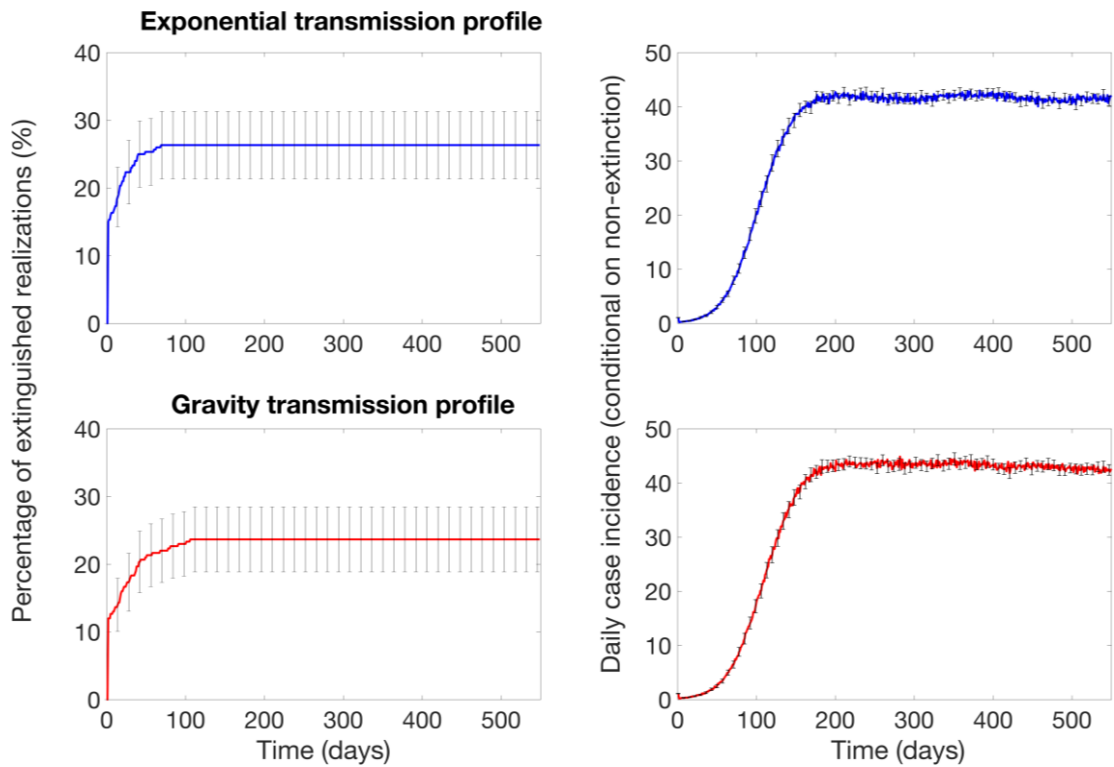

**Figure 5.** The impact of ring vaccination that starts 30 days after epidemic onset on the probability of outbreak extinction and the mean daily incidence curves for various percentage levels of inaccessible households, which do not participate in vaccination and do not provide contacts lists as explained in the text. These simulations were generated using the exponential transmission profile, but similar results were obtained using the gravity transmission profile described in the main text. The vertical dashed line indicates the timing of start of the ring vaccination program.

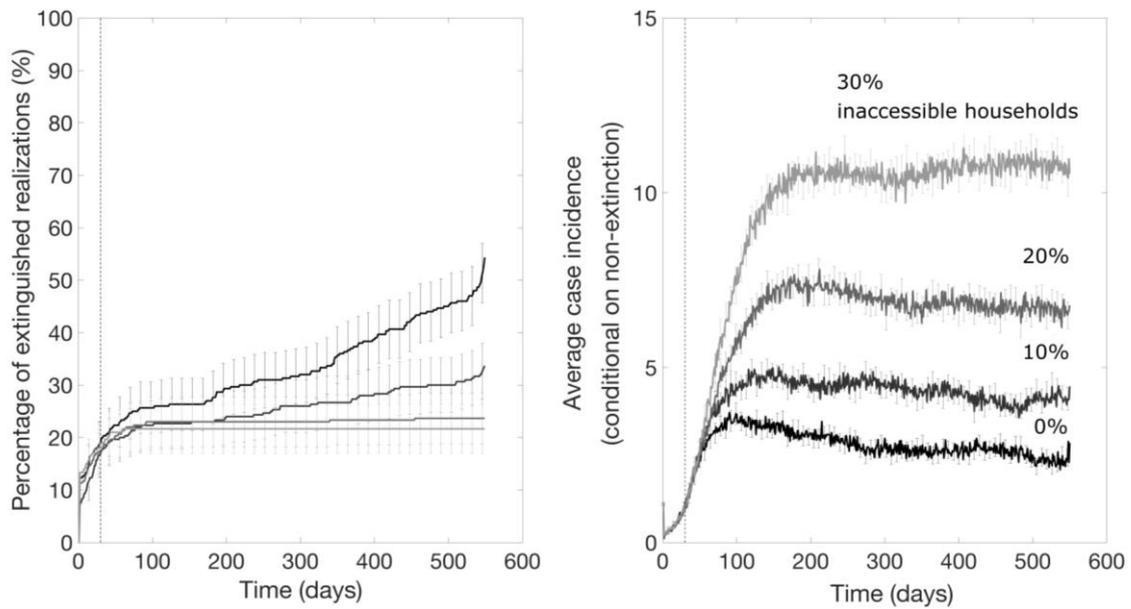

**Figure 6.** The probability of epidemic control for a ring vaccination strategy as a function of the radius of the ring and the delay to vaccinating contacts. The baseline parameter value for the radius of ring vaccination is the intermediate value of $R_{VR}=5$, and the baseline parameter value for the delay to vaccination is the intermediate value of $\tau=6$ days. Baseline parameter values for other parameters not varied in this figure are given in Table 1.

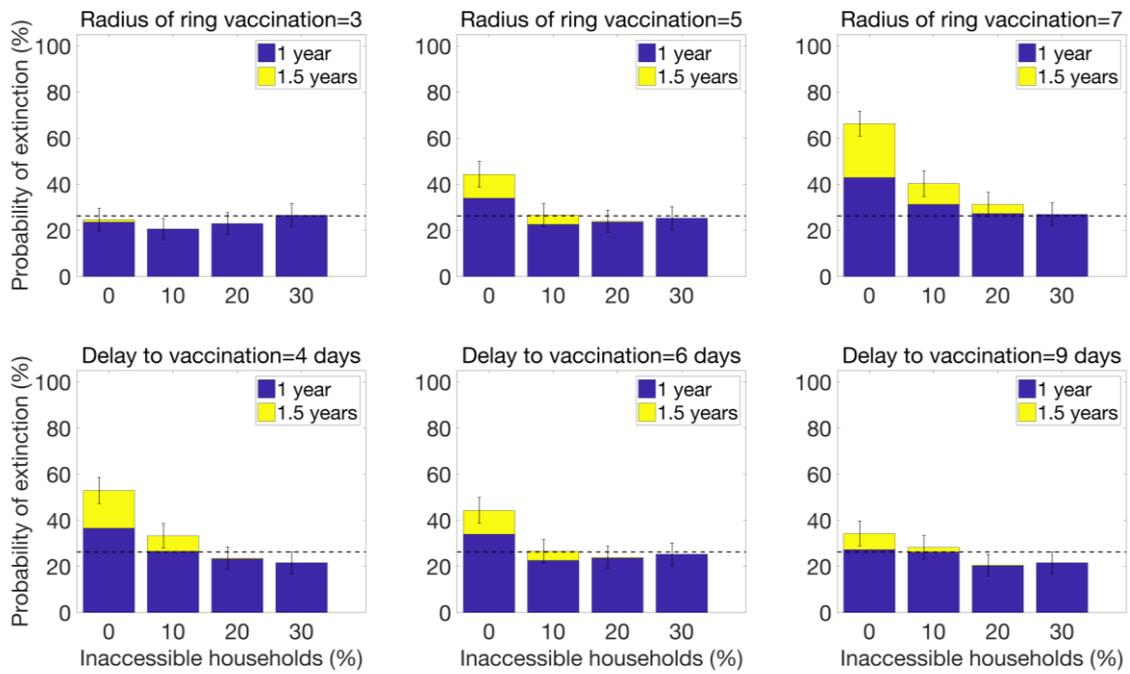

**Figure 7.** The impact of ring vaccination on an established epidemic wave (after a delay of 270 days) on the probability of outbreak extinction and the mean daily incidence curves for various percentage levels of household inaccessibility. These simulations were generated using the exponential transmission profile, but similar results were obtained using the gravity transmission profile described in the main text. The vertical dashed line indicates the timing of start of the ring vaccination program.

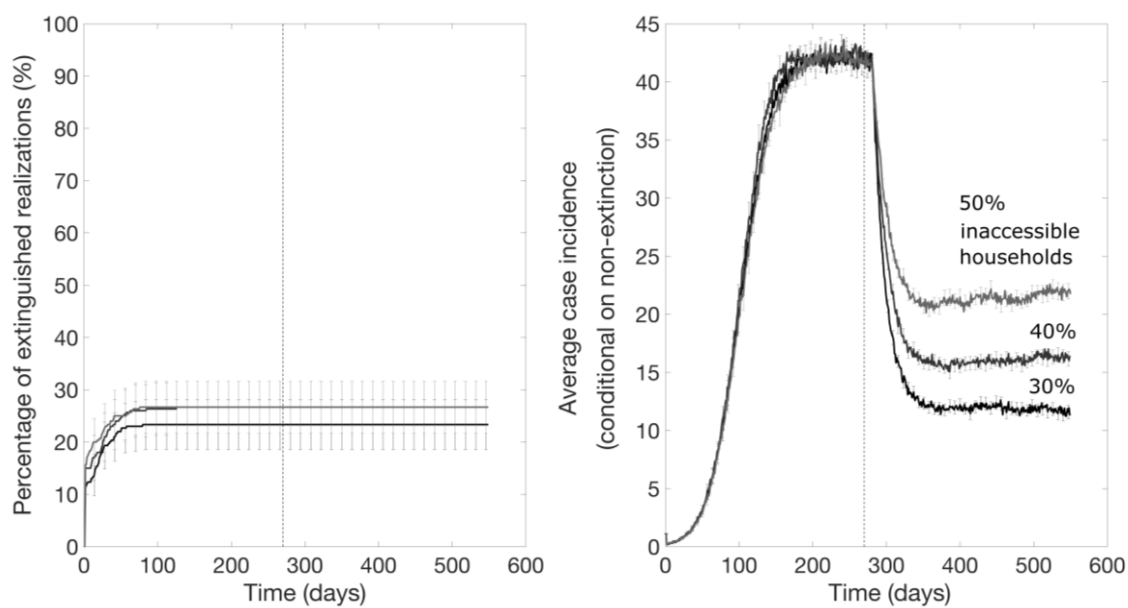

**Figure 8**. Mean daily case incidence when a community vaccination rate of 10% per day supplements a ring vaccination strategy 9 months later for different levels of household inaccessibility. The vertical dashed line indicates the timing of start of the supplemental community vaccination efforts. These simulations were generated using the exponential transmission profile, but similar results were obtained using the gravity transmission profile described in the main text.

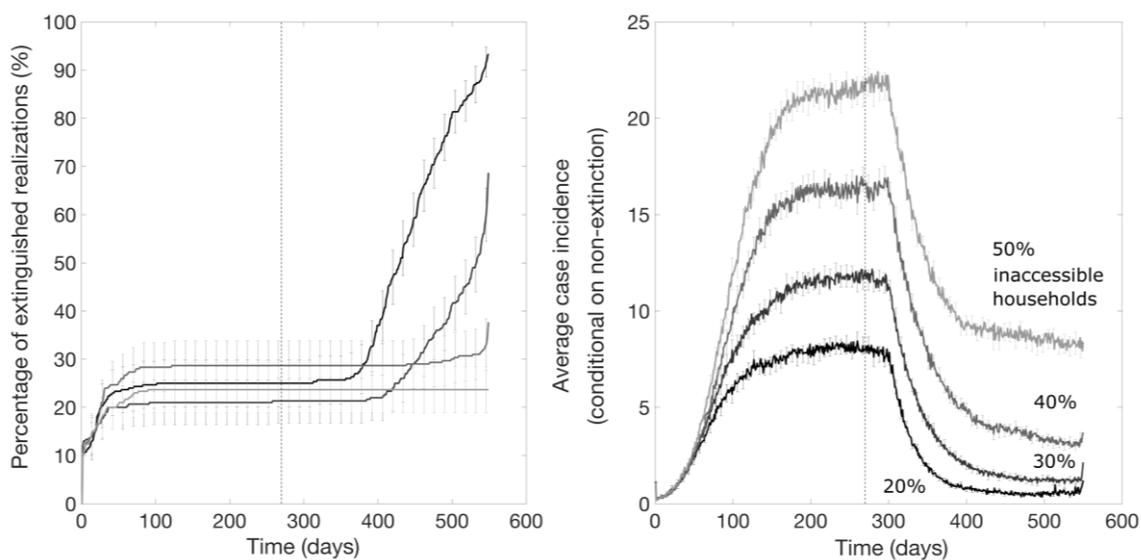

**Figure 9**. Probability of epidemic extinction over time and the mean daily case incidence when different community vaccination rates supplement a ring vaccination strategy for a 30% level of household inaccessibility and community vaccination starts 9 months after epidemic onset. The vertical dashed line indicates the timing of start of community vaccination. These simulations were generated using the exponential transmission profile, but similar results were obtained using the gravity transmission profile described in the main text.

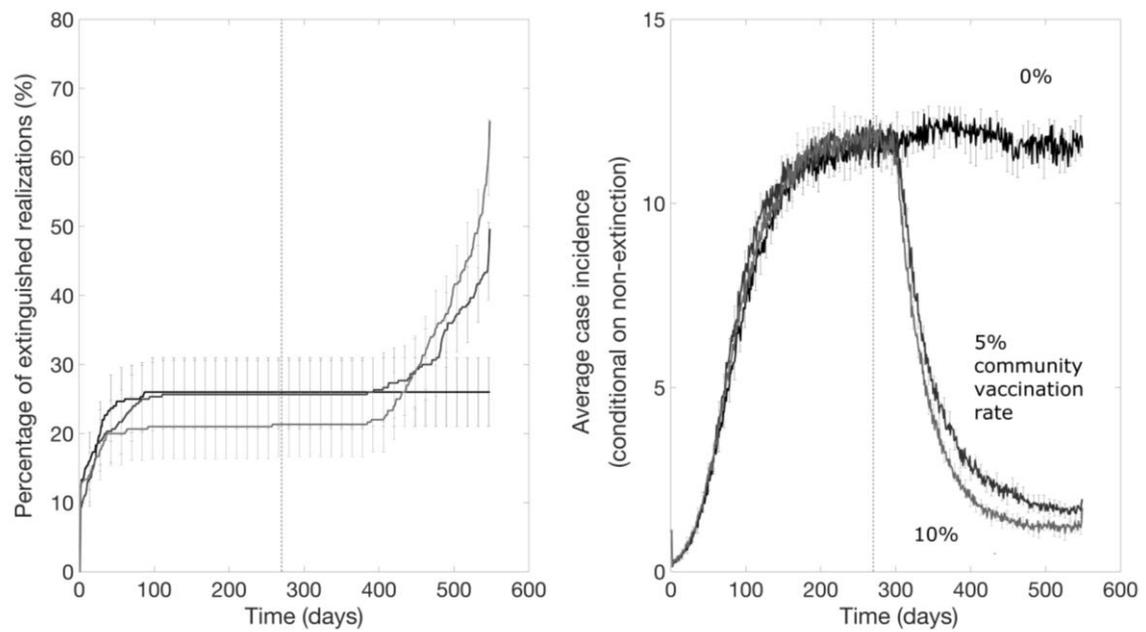

**Figure 10**. Supplemental effect on community vaccination on the probability of epidemic extinction for populations with different household inaccessibility levels ranging from 0 to 50%. Bars show the probability of epidemic extinction after 1.5 years for increasing vaccination measures: i) ring vaccination alone (blue bars), ii) ring vaccination with supplemental community vaccination with a rate of 5% community vaccination per day (light blue bars) and iii) ring vaccination with supplemental community vaccination with a rate of 10% vaccination per day. The horizontal line indicates the mean probability of extinction for the baseline scenario in the absence of vaccination. Timing was chosen to resemble the scenario for the DRC Ebola epidemic, with ring vaccination applied after 7 days and community vaccination applied after 9 months. These simulations were generated using the exponential transmission profile, but qualitatively similar results were obtained using the gravity transmission profile described in the main text.

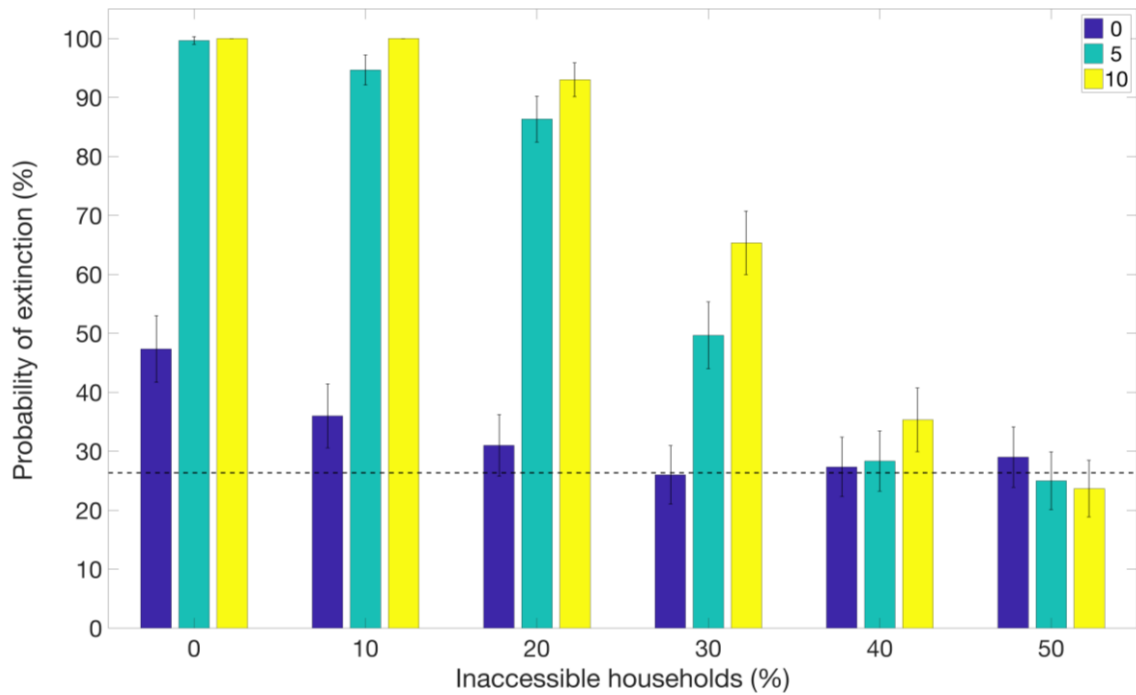